\documentstyle[twoside,fleqn,espcrc2,epsfig]{article}

\def\bef{\begin{figure}}
\def\eef{\end{figure}}
\newcommand{\be}[1]{\begin{equation}\label{#1}}
\newcommand{\beq}{\begin{equation}}
\newcommand{\ee}{\end{equation}}
\newcommand{\beqn}[1]{\begin{eqnarray}\label{#1}}
\newcommand{\eeqn}{\end{eqnarray}}
\newcommand{\bd}{\begin{displaymath}}
\newcommand{\ed}{\end{displaymath}}

\newcommand{\matr}[9]{\left(\begin{array}{ccc}{#1}&{#2}&{#3}\\
{#4}&{#5}&{#6}\\{#7}&{#8}&{#9}\end{array}\right)}
\def\lsim{\raise0.3ex\hbox{$\;<$\kern-0.75em\raise-1.1ex
\hbox{$\sim\;$}}}
\def\gsim{\raise0.3ex\hbox{$\;>$\kern-0.75em\raise-1.1ex
\hbox{$\sim\;$}}}
\def\simlt{\mathrel{\lower2.5pt\vbox{\lineskip=0pt\baselineskip=0pt
           \hbox{$<$}\hbox{$\sim$}}}}
\def\simgt{\mathrel{\lower2.5pt\vbox{\lineskip=0pt\baselineskip=0pt
           \hbox{$>$}\hbox{$\sim$}}}}
\def\unity{{\hbox{1\kern-.8mm l}}}
\def\16p{16\pi^2}

\def\dem{\delta m^2}

\def\bY{{\bf Y}}

\def\bG{{\bf G}}

\newcommand{\AmS}{{\protect\the\textfont2
  A\kern-.1667em\lower.5ex\hbox{M}\kern-.125emS}}

\title{ {
\vskip -0.2cm 
\hfill{\small DFAQ/99/TH/01}\\
\vskip -0.25cm 
\hfill{\small DFPD/99/TH/28}}
\vskip 0.7cm
Towards a grand  
unified picture for neutrino and quark mixings
\thanks{Talk given by A. Rossi 
at the Int. Workshop ``Particles in Astrophysics 
and Cosmology: from Theory to Observation'', 
May 3-8, 1999, Valencia, Spain.}
\author{Zurab Berezhiani 
\address{
Universit\`a dell'Aquila, 
I-67010 Coppito, L'Aquila, Italy, and \\
Institute of Physics, Georgian Academy of Sciences,
380077 Tbilisi, Georgia} and
Anna Rossi\address{ Universit\`a di Padova and 
INFN Sezione di Padova, I-35131 Padova, Italy
}}
}
\begin{document}

\begin{abstract}
The comparison of the CKM mixing angles with the leptonic mixings implied 
by the recent atmospheric and solar neutrino data 
exhibits an interesting complementarity.
This pattern can be understood in the context of 
the $SU(5)$ grand unification, assuming that the fermion 
mass matrices have  Fritzsch-like structures 
but  are not necessarily symmetric.
(The present contribution is based on the paper in ref. \cite{az}.)

\end{abstract}

\maketitle

\section{Introduction}
One of the mysteries of  particle physics 
is the manifest hierarchy in the fermion spectrum and mixing angles. 
The masses of the quarks and charged leptons are spread over
five orders of magnitude, from MeVs to hundreds of GeVs and  
the quark mixing angles are:
\beqn{q-ang}
\theta_{23}^q& = &(2.3 \pm 0.2)^\circ, ~~~~ 
\theta_{12}^q = (12.7 \pm 0.1)^\circ, \nonumber \\ 
\theta_{13}^q &= &(0.18 \pm 0.04)^\circ 
\eeqn

As for the neutrinos, the recent data from the atmospheric and 
solar neutrino (AN and SN) experiments 
\cite{rev} providing  information on their masses and mixings, 
have made the mystery of  ``flavour'' even  more intriguing.
On the one hand, the ranges of 
$\dem_{\rm atm}$ and $\dem_{\rm sol}$
needed for the explanation of the AN and SN anomalies, 
can be translated directly into  values of the neutrino masses. 
Namely, assuming the mass hierarchy $m_3 >m_2> m_1$ for the 
neutrino mass eigenstates $\nu_{1,2,3}$ we find
a mass hierarchy $m_2/m_3$ similar to that of the charged 
leptons:\footnote{
Below we concentrate on the small-mixing angle MSW solution 
for the SN problem \cite{MSW}, barring other possibilities such as  
the large-mixing angle MSW or vacuum oscillation solutions.}
\be{nu-spectr} 
m_3\!\! =\! (5.7_{-2.2}^{+2.7})\cdot 10^{-2} ~ {\rm eV}, ~~
m_2\!\! = \!(2.5_{-0.5}^{+0.7})\cdot 10^{-3} ~ {\rm eV} 
\ee
On the other hand, 
the magnitudes of the neutrino mixing 
angles\footnote{
For $\dem_{\rm atm} > 2\cdot 10^{-3}$ eV$^2$ 
the limit  $\theta^l_{13}<13^\circ$ follows 
from the CHOOZ experiment.
Moreover, taking into account all the experimental data, 
$\theta^l_{13}\approx 0$ provides the best data fit 
both for AN and SN cases \cite{BHSSW}. 
}  
\beqn{nu-ang}
\theta_{23}^l&=& (45\pm 11)^\circ , ~~~
\theta_{12}^l= (2.0\pm 1.2)^\circ , \nonumber \\
\theta_{13}^l& < &(13-20)^\circ  
\eeqn 
are in clear contrast with the corresponding quark angles 
(\ref{q-ang}).   
In short: the AN anomaly points to  maximal 23 mixing 
in the leptonic sector 
to be compared with the  very small 23 mixing of quarks, 
and on the contrary, the MSW solution implies a very small 12 
lepton mixing angle  versus the reasonably large value 
of the Cabibbo angle. 

In the standard model (SM) or in its supersymmetric 
extension the masses of the charged fermions
$q_i=(u_i,d_i)$, $u^c_i$, $d^c_i$; 
$l_i=(\nu_i,e_i)$, $e^c_i$  ($i=1,2,3$ is a family index)  
emerge from the Yukawa terms: 
\be{Yuk}
\phi_2 u^c_i \bY_u^{ij} q_j 
+ \phi_1 d^c_i \bY_d^{ij} q_j + \phi_1 e^c_i \bY_e^{ij} l_j 
\ee
where $\phi_{1,2}$ are the Higgs doublets: 
$\langle \phi_{1,2} \rangle = v_{1,2}$, 
$(v_1^2 + v_2^2)^{1/2} =v_w=174$ GeV
and $\bY_{u,d,e}$ are arbitrary matrices 
of  coupling constants.
The neutrino masses emerge only from the higher 
order effective operator \cite{BEG}: 
\be{Yuk-nu}
\frac{\phi_2\phi_2}{M}\, l_i \bY_\nu^{ij} l_j \,, ~~~~~~~ 
\bY_\nu^{ij}=\bY_\nu^{ji} 
\ee
where $M\gg v_w$ is some cutoff scale and  
$\bY_\nu$ is a 
matrix of  dimensionless coupling constants.   
The fermion mass eigenstates are identified by diagonalizing 
the Yukawa matrices $\bY_{u,d,e,\nu}$  
by  bi-unitary transformations: 
\be{diag} 
U'^T_{f}\bY_f U_{f}=\bY_f^{D}\,, ~~~ f=u,d,e,\nu
\ee
(for the neutrinos it is $U'_{\nu }\equiv U_{\nu}$). 
In this way  
the Cabibbo-Kobayashi-Maskawa (CKM) matrix 
$V_q=U^\dagger_{u}U_{d}$ and the leptonic 
mixing matrix ${V}_l=U_{e}^\dagger U_{\nu}$,
describing  the neutrino oscillation phenomena,   
are also determined: 
\beqn{CKM}
&&V_q = \matr{V_{ud}}{V_{us}}{V_{ub}} {V_{cd}}{V_{cs}}{V_{cb}} 
{V_{td}}{V_{ts}}{V_{tb}} ~,\\
&&{V}_l =  
\matr{V_{e1}}{V_{e2}}{V_{e3}} {V_{\mu1}}{V_{\mu2}}{V_{\mu3}} 
{V_{\tau1}}{V_{\tau2}}{V_{\tau3}} 
\eeqn
%
For both  mixing matrices, 
we adopt the ``standard''  parametrization utilizing  
the angles $\theta_{12},~\theta_{23},\theta_{13}$ ~ and a 
CP-phase $\delta$.\footnote{
The leptonic mixing matrix contains two additional phases 
that are not relevant for the neutrino oscillations.}
In the following, we distinguish 
the quark and lepton mixing angles in $V_q$ and $V_l$ 
by the subscripts `$q$' and `$l$', respectively.

As already mentioned, the SM does not provide any theoretical 
hints to constrain the matrices $\bY_{u,d,e}$ and $\bY_\nu$, leaving 
the issue of the  
fermion mass hierarchy and mixing pattern  unexplained. 
Concerning the neutrinos, also the mass scale $M$ remains  a free 
parameter. One can only conclude that if the maximal 
constant in $\bY_\nu$ is of order the top Yukawa constant,  
$Y_3\sim Y_t\sim 1$, then the mass value $m_3$ in 
(\ref{nu-spectr}) points to the scale $M\sim 10^{15}$ GeV, 
rather close to the grand unified scale.  

In this respect, the grand unified theories can be very useful.
In these theories, as a consequence 
of the larger gauge group, relationships between quark and lepton 
 masses or between  CKM angles and quark mass ratios
can emerge naturally. 
Moreover, the assumption of further symmetries 
in the Yukawa sector --- well known examples being the 
``horizontal'' or ``family'' symmetries --- implies 
further predictions and thus a possible clue to discern 
the ``flavour'' mystery \cite{all,GUT}. 

A popular Yukawa texture 
is that suggested by Fritzsch \cite{Fritzsch}:
\be{Fr}
\bY_{u,d,e}=\,\matr{0}{A'_{u,d,e}}{0}
{A_{u,d,e}}{0}{B'_{u,d,e}}{0}{B_{u,d,e}}{C_{u,d,e}}
\ee 
where all elements are generically  complex and obey 
the additional condition:  
\be{A-B}
|A'_f|=|A_f|, ~~~~ |B'_f|=|B_f|; ~~~~~ f=u,d,e 
\ee
The presence of zero elements as well as the 
``symmetricity'' property (\ref{A-B}) 
can be motivated by `` horizontal''  
symmetries \cite{PLB85}. 
This pattern has many interesting properties. 
For instance,  it  links 
the observed value of the Cabibbo angle, 
$V_{us}\approx \sqrt{m_d/m_s}$,  
to the observed size of the CP-violation in the $K-\bar{K}$ system 
and the predicted magnitude 
$|V_{ub}/V_{cb}|\approx \sqrt{m_u/m_c}$ is  in 
good agreement with the data.  
Unfortunately, this texture implies cannot account at the same time 
for   the small value of  $V_{cb}$ and 
the large top mass.

However, this shortcoming can be cured  just 
by {\it embedding} the ansatz in a $SU(5)$ grand unified theory and 
{\it breaking} the symmetricity condition\footnote{
The need of such an asymmetry was invoked in the context of $SO(10)$ models 
\cite{GUT}.} in the 23-family sector 
with $b_e=\left|B_e/B'_e\right| > ~1$ and 
$b_d=\left|B'_d/B_d \right|> ~1$.  
The $SU(5)$ symmetry ensures the  following  product rule for the 
mixing angles:
\be{rule23} 
\tan\theta_{23}^d \tan\theta_{23}^e \sim 
\left(\frac{m_\mu m_s}{m_\tau m_b}\right)^{1/2}
\ee
This rule is certainly exact when 
the down-quark and charged-lepton matrices 
have the symmetric Fritzsch texture from which one derives   
$\tan\theta_{23}^d= (m_s/m_b)^{1/2}$ and 
$\tan\theta_{23}^e = (m_\mu/m_\tau)^{1/2}$.   
However, these two relations are unsatisfactory as
$|V_{cb}|< (m_s/m_b)^{1/2}$ and  
$\sin\theta_{atm} < (m_\mu/m_\tau)^{1/2}$.
On the other hand, whenever 
the symmetricity condition is broken,  
the rule (\ref{rule23}) is only approximate since 
none of those angles can be predicted in terms of mass 
ratios. Indeed their values now depend on the amount of asymmetry 
between the 23 and 32 entries, i.e. on the  factors $b_d$ and $b_e$.
One can easily realize that the 
increasing of $b_e$ goes in  parallel 
with that of $b_d$   
since in SU(5) the Yukawa matrices are related 
as $\bY_e = \bY_d^T$, modulo certain Clebsch factors. 
As a result the 23 mixing   becomes larger 
in the leptonic sector and   smaller  in the quark sector. 
Therefore, if $\tan\theta_{23}^d$ decreases below  
$(m_s/m_b)^{1/2}$, then  $\tan\theta_{23}^e$ should 
correspondingly increase above $(m_\mu/m_\tau)^{1/2}$, 
and when the former reaches the value $|V_{cb}|\simeq 0.05$, 
the latter becomes $\sim 1$  
(this happens for $b_{d,e}\sim 8$). 
Though these estimates are not precise,  
they qualitatively 
demonstrate the `seesaw' correspondence between the 
quark and lepton mixing angles whenever their magnitudes are 
dominated by the rotation angles coming from the 
down fermions.
A similar argument can be applied also to the 12 mixing: 
\be{rule12}  
\tan\theta_{12}^d \tan\theta_{12}^e \sim 
\left(\frac{m_e m_d}{m_\mu m_s}\right)^{1/2}
\ee
The relation $V_{us}\simeq (m_d/m_s)^{1/2}$ 
suggests that the 12 block of $\bY_d$ should be nearly symmetric, 
and hence we expect that $\sin\theta_{sol}\sim (m_e/m_\mu)^{1/2}$.

The above discussion is the key-point that will be extensively 
developed and discussed in the next section.
 
\section{Modifying the Fritzsch ansatz in $SU(5)$}
In the $SU(5)$ model the masses of the fermions  
$\bar{5}_i \!=\!(d^c, l)_i$,  $10_i \!=\!(u^c, e^c, q)_i$   
arise from the following Yukawa terms:
\be{Yuk-su5}
\bar{H} 10_i\bG^{ij}\bar5_j  +  H 10_i\bG_u^{ij}10_j + 
\frac{HH}{M}\bar5_i\bG_\nu^{ij}\bar5_j 
\ee
where $H=(T,\phi_2)\sim 5$ and $\bar H=(\bar T,\phi_1)\sim \bar5$\\ 
are the Higgses. 
The Yukawa constant matrices $\bG_{u}$ and $\bG_{\nu}$ 
are symmetric due to $SU(5)$ symmetry reasons 
while the form of $\bG$ is not constrained. 
Upon breaking the $SU(5)$ symmetry, we recover the SM Yukawa couplings 
(\ref{Yuk}) with 
\be{relat}
\bY_e=\bG, ~~~ \bY_d=\bG^T, ~~~ \bY_u=\bG_u, ~~~
\bY_\nu=\bG_\nu
\ee
To simplify the discussion we shall assume, without loss of generality, 
that the matrices $\bG_u$ and $\bG_\nu$ are diagonal.   
Then the weak mixing 
matrices in  the quark and leptonic sectors are just 
$V_q=U_d$ and $V_l=U_e^\dagger$. 
On the other hand, 
since $\bY_d=\bY_e^T$, 
we get that $U_{d}=U'_{e}$ and $U_{e}=U'_{d}$, 
so that the rotation angles of the left down quarks 
(charged leptons) are related to the unphysical angles 
rotating the right states of the charged leptons 
(down quarks).
In the minimal $SU(5)$ model the entries of the matrix $\bG$ are just 
constants and one faces the well-known problem of the 
down-quark and charged-lepton degeneracy at the GUT scale. 
While  the $Y_b=Y_\tau$ unification  is a success 
of the SUSY $SU(5)$ GUT, 
the other predictions 
$Y_{s,d}=Y_{\mu,e}$ are clearly wrong.

A more satisfactory picture emerges if the terms  
$\bar{H}10_i\bG^{ij}\bar5_j$ are understood as effective cubic couplings 
originating from 
higher-order operators, such as  
$\bar{H}10_i(\frac{\Phi}{M_s}\hat{G}_{ij})
\bar5_j$,   where $\Phi$ is the 
$SU(5)$ adjoint and $M_s$ is some fundamental scale larger than 
the GUT scale. As a consequence, the corresponding entries in $\bY_e$ and 
 $\bY_d$ can be distinguished by Clebsch coefficients.

\begin{figure}[t]
\vskip -1.0cm
\centerline{\protect\hbox{\epsfig{file=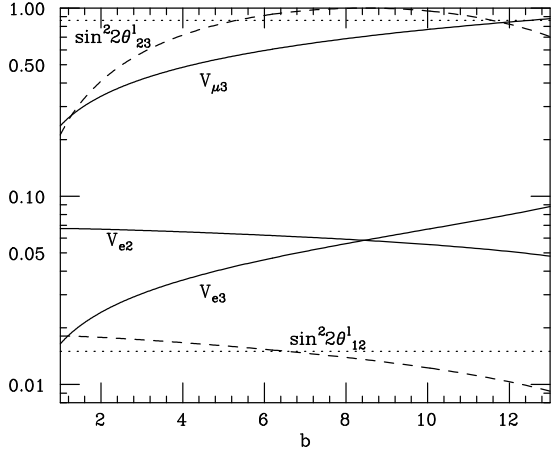,height=9.6cm,
width= 7.5cm,angle=90}}}
\vskip -1.7cm
\caption{The lepton mixing angles (symbols on the curves)  
as functions of $b$ (solid). 
The oscillation parameters 
$\sin^22\theta^l_{23}$ and $\sin^22\theta^l_{12}$  
are also shown (dash), together with  the experimental 
limits $\sin^22\theta^l_{23} >0.86$ and 
$\sin^22\theta^l_{12}< 1.5 \cdot10^{-2}$ (dotted lines). 
}
\vskip -0.8cm
\label{f1}
\end{figure}

In this perspective the matrices $\bY_e$ and $\bY_d$ can assume the 
asymmetric form 
given in Eq.~(\ref{Fr}). 
Phenomenological arguments impose these further relations:
\beqn{restrict} 
&&
C_d=C_e ~(=C),  \nonumber \\ 
&&
A_d=A'_d=A'_e=A_e ~(=A), \nonumber\\
&&
 B'_d = k'B_e,  ~~~~~~ B_d = k B'_e  
\eeqn 
where the coefficients $k$ and $k'$ are 
 nontrivial $SU(5)$ Clebsches 
breaking the quark and lepton symmetry.
Introducing the 23-sector asymmetry parameters $b_e=B_e/B'_e$ 
and $b_d=B'_d/B_d = \frac{k'}{k}b_e$ we finally end up with the following 
textures:
\beqn{Fred}
&&\bY_e \!= \!\matr{0}{A}{0} {A}{0}{\frac{1}{b}B} {0}{B}{C}, \\
&&\bY_d \!= \!\matr{0}{A}{0} {A}{0}{k'B} {0}{\frac{k}{b}B}{C}
\eeqn
This ansatz depends on six parameters:
three Yukawa entries $A,B,C$ and three Clebsch factors 
$k,k'$ and $b$. Through these parameters 
we have to determine six eigenvalues --
$Y_{e,\mu,\tau}$ and $Y_{d,s,b}$ -- and  six mixing angles --
$s^q_{12},s^q_{23},s^q_{13}$ and 
$s^l_{12},s^l_{23},s^l_{13}$.
Hence at the GUT scale we are left 
with six relations between the physical observables. 
The leptonic mixing angles can be then expressed in terms of  ratios 
of the corresponding lepton masses and the asymmetry parameter $b_e$. 
Fig.~\ref{f1} illustrates  the $b$-dependence of  
the leptonic mixing angles and of the corresponding  parameters 
 $\sin^2 2\theta^l_{23} = 4|V_{\mu3}|^2(1-|V_{\mu3}|^2)$ 
and $\sin^2 2\theta^l_{12} = 4|V_{e2}|^2(1-|V_{e2}|^2)$.

\begin{figure}[htb]
\vskip -1.0cm
\centerline{\protect\hbox{\epsfig{file=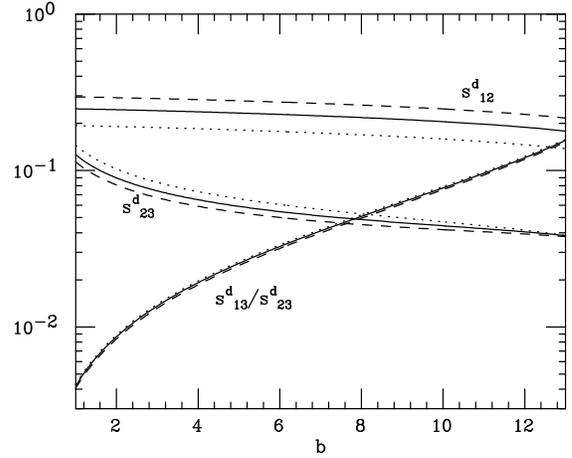,height=9.6cm,width= 7.5cm,angle=90}}}
\vskip -1.7cm
\caption{As in Fig. 1 for the quark mixing angles, with   
$b_d=b$ (i.e. $k'=k$), for 
three different values:  $k^2=1/4$ (solid),
$k^2=1/3$ (dott) and $k^2=1/5$ (dash).
All mixing angles  are evaluated at the GUT scale.
}
\label{f2}
\vskip -0.8cm
\end{figure}

\begin{figure}[htb]
\vskip -1.0cm
\centerline{\protect\hbox{\epsfig{file=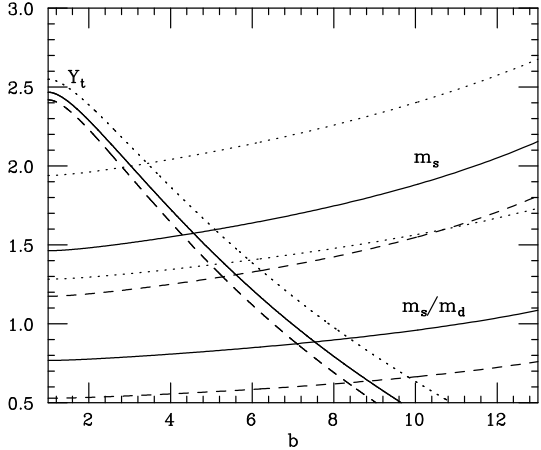,height=9.6cm,width= 7.5cm,angle=90}}}
\vskip -1.8cm
\caption{The dependence of the down quark masses on $b_e=b$.
$m_s(1~{\rm GeV})$ is shown in units of 100 MeV and the
 ratio $m_s/m_d$ in units of 20. 
We also show iso-contours for $m_b (m_b)=4.25$ GeV 
in the $b-Y_t$ plane (for $\alpha_s=0.118$). 
}
\label{f3}
\vskip -0.8cm
\end{figure}
For $b=1$ the 23 mixing  angle 
 is rather small for 
explaining the AN anomaly,  while the 12 mixing 
is somewhat above the upper limit 
obtained by the MSW fit of the SN data (c.f. (\ref{nu-ang})). 
However, for larger $b$, $|V_{\mu3}|$ increases roughly as 
$\sqrt{b}$ and becomes maximal around $b=8.4$, 
while $|V_{e2}|$ slowly decreases 
(roughly as $\sqrt{c^e_{23}}$). 
Thus, the AN bound, $\sin^2 2\theta_{23} >0.86$,  
requires $6<b<12$, while  the  SN data favour  
$b>7$, when $\sin^2 2\theta^l_{12}$ 
drops below $1.5\cdot 10^{-2}$.

Analogously the quark masses and mixing angles can be expressed in terms 
of the lepton mass ratios and of the three parameters $b_e, k,k'$. 
Then we  show the 
behaviour  of  the mixings (Fig.~\ref{f2}),  
of the masses\footnote{The re-normalization scaling has been taken 
into account for the bottom mass.} $m_s$, $m_b$ and the ratio 
$m_s/m_d$ (Fig.~\ref{f3}) for several values of 
$k\cdot k'$ .  
For $k\sim k'$ and large values of $b$, 
($b=7-12$ as required from the lepton mixing) we achieve
quite a satisfactory description  also of the 
quark sector.
The pattern with $k=k'=1/2$ looks somehow favoured. 
We also learn from 
Fig.~\ref{f3} that rather small values of
$Y_t\sim 0.5-1$ are needed to obtain the 
correct bottom  mass for $b\gsim 7$.
This  special feature  arises  from the
substantial correction to the $b-\tau$ 
Yukawa unification due to the large $b$.

In a more general case, we have to expect  also $\bY_u,~\bY_\nu$ to have 
a Fritzsch-like form. This would occur in the presence of some underlying 
horizontal symmetry.   
Such a scenario would provide some different features. 
In this case smaller values of $b_{e,d}$ can suffice 
since now the mixing angles will be contributed also 
by the unitary matrices $U_u$ and $U_\nu$: 
$V_q=U_u^\dagger U_d$ and $V_l=U_e^\dagger U_\nu$. 
For the CKM mixing angles we have: 
\beqn{Vcb}
&&|V_{cb}|=s^q_{23}\approx  
\left|s^d_{23}-e^{i\varphi}s^u_{23}\right|, 
\nonumber \\ 
&&|V_{us}|=s^q_{12} \approx
\left|s^d_{12}-e^{i\delta}s^u_{12}\right|, 
\left|\frac{V_{ub}}{V_{cb}}\right|\approx s^u_{12} 
\eeqn
where the phases $\varphi$, $\delta$ etc are combinations 
of the independent phases in the Yukawa matrices. 
The $\theta^u_{23}$, $\theta^u_{12}$ are the analogous angles 
diagonalizing $\bY_u$:  
$\tan\theta^u_{23}=\sqrt{Y_c/Y_t}$ and 
$\tan\theta^u_{12}=\sqrt{m_u/m_c}$. 
By varying  the phase $\varphi$  from $0$ to $\pi$, 
the value of the 23 mixing angle in the CKM matrix can 
vary between its minimal and maximal possible values:  
\be{pm}
\theta^{q(\mp)}_{23} = \theta^d_{23} \mp \theta^u_{23} 
\ee
Analogously, for the leptonic mixing we have 
\be{pm-nu23}
\theta^{l(\mp)}_{23} = \theta^e_{23} \mp \theta^\nu_{23} 
\ee
where $\tan\theta^\nu_{23}= \sqrt{m_2/m_3}$. Thus, 
for the range of the neutrino masses indicated in 
(\ref{nu-spectr}) we obtain 
$\theta_{23}^\nu= (11.8_{-3.0}^{+5.0})^\circ$.
In case of  moderate asymmetry in $\bY_{d,e}$, 
the entries in (\ref{pm}) are big as compared to the 
experimental value of $\theta^{q}_{23}$ while each 
of the entries in (\ref{pm-nu23}) is too small for the 
magnitude of $\theta^{l}_{23}$ required by the AN oscillation. 
However, by properly tuning the phases, 
 $\theta^{q}_{23}$  can get close to 
$\theta^{q(-)}_{23}=\theta^d_{23} - \theta^u_{23} $ 
while $\theta^{l}_{23}$ can approach 
$\theta^{l(+)}_{23}=\theta^e_{23} + \theta^\nu_{23} $,  
Therefore, even for small values $b_{e,d}\approx 2$, one could 
achieve a proper fit of the mixing angles.
In ref. \cite{az} an example of realization of such a scenario,
implementing the $U(2)$ horizontal symmetry,  is illustrated.  
 
\section{Conclusions}
We have discussed how the present pattern of the 
leptonic mixing angles, characterized by a maximal mixing between 
the second and third generation, can be linked to the CKM mixing angles 
in the $SU(5)$ grand unification thanks  to the fermion multiplet structure. 
In particular, this has been shown assuming the fermion Yukawa matrices 
to have a Fritzsch-like form with an asymmetric 23-block and (essentially) 
symmetric 12-block.
 
We remark that  alternative  and realistic  ans\"atze 
-- with diagonal $\bY_{u,\nu}$  -- (accounting e.g. 
for  CP-violation) 
can be motivated in the context of $U(3)$ horizontal symmetry \cite{za}.

\section{Acknowledgements}

A. R.  wishes to thank   Jose Valle 
and all organizers  of the Conference 
for the pleasant and
interesting atmosphere.

\end{document}